\documentclass[aps,prd,twocolumn,nofootinbib,superscriptaddress,preprintnumbers,floatfix]{revtex4}

\usepackage{graphicx}
\usepackage[hypertex]{hyperref}

\newcommand{\nn}{\nonumber}
\newcommand{\be}{\begin{equation}}
\newcommand{\ee}{\end{equation}}
\newcommand{\beqa}{\begin{eqnarray}}
\newcommand{\eeqa}{\end{eqnarray}}

\newcommand{\GeV}{~\text{GeV}}
\newcommand{\TeV}{~\text{TeV}}
\newcommand{\MeV}{~\text{MeV}}

\newcommand{\Br}{\text{Br}}

\newcommand{\Sec}[1]{Sec.~\ref{#1}}

\newcommand{\Fig}[1]{Fig.~\ref{#1}}
\newcommand{\Ref}[1]{Ref.~\cite{#1}}
\newcommand{\Eq}[1]{Eq.~(\ref{#1})}

\def\d{{\rm d}}

\makeatletter
\def\simgt{\mathrel{\lower2.5pt\vbox{\lineskip=0pt\baselineskip=0pt
      \hbox{$>$}\hbox{$\sim$}}}}
\def\simlt{\mathrel{\lower2.5pt\vbox{\lineskip=0pt\baselineskip=0pt
      \hbox{$<$}\hbox{$\sim$}}}}
\makeatother

\begin{document}

\title{\boldmath Constraining the Axion Portal with $B \to K\ell^+\ell^-$}

\author{Marat Freytsis}
\affiliation{Berkeley Center for Theoretical Physics, Department of Physics,
  University of California, Berkeley, CA 94720}
\affiliation{Theoretical Physics Group, Lawrence Berkeley National Laboratory,
  University of California, Berkeley, CA 94720}

\author{Zoltan Ligeti}
\affiliation{Theoretical Physics Group, Lawrence Berkeley National Laboratory,
  University of California, Berkeley, CA 94720}

\author{Jesse Thaler}
\affiliation{Berkeley Center for Theoretical Physics, Department of Physics,
  University of California, Berkeley, CA 94720}
\affiliation{Theoretical Physics Group, Lawrence Berkeley National Laboratory,
  University of California, Berkeley, CA 94720}

\begin{abstract}
We investigate the bounds on axionlike states from flavor-changing neutral
current $b\to s$ decays, assuming the axion couples to the standard model
through mixing with the Higgs sector.  Such GeV-scale axions have received
renewed attention in connection with observed cosmic ray excesses.  We find that
existing $B\to K\ell^+\ell^-$ data impose stringent bounds on the axion decay
constant in the multi-TeV range, relevant for constraining the ``axion portal''
model of dark matter.  Such bounds also constrain light Higgs scenarios in the
next-to-minimal supersymmetric standard model. These bounds can be improved by
dedicated searches in $B$-factory data and at LHCb.

\end{abstract}

\maketitle

\section{Introduction}
\label{sec:intro}

Motivated by a variety of cosmic ray anomalies~\cite{Adriani:2008zr,
Abdo:2009zk,Collaboration:2008aaa,Aharonian:2009ah}, a new dark matter paradigm
has emerged where TeV-scale dark matter interacts with GeV-scale bosons
\cite{Finkbeiner:2007kk, Pospelov:2007mp,ArkaniHamed:2008qn}.  In one such
scenario --- dubbed the ``axion portal"~\cite{Nomura:2008ru} --- dark matter in
the Milky Way halo annihilates into light pseudoscalar ``axions''.  In order to
explain the observed galactic electron/positron excess, the axion, $a$, is
predicted to have a specific mass and decay constant~\cite{Nomura:2008ru}
\be\label{jyrange}
360 < m_a \lesssim 800 \MeV, \qquad f_a \sim 1-3 \TeV.
\ee
These axions couple to standard model fermions proportional to their Yukawa
couplings, and in this mass range the axion dominantly decays as $a\rightarrow
\mu^+\mu^-$.  Other novel dark matter scenarios involving axionlike states have
also been
proposed~\cite{Ibe:2009dx,Bai:2009ka,Mardon:2009gw,Hooper:2009gm,Ibe:2009en},
which allow for broader range of axion masses and decay constants.

More generally, light axionlike states appear in a variety of new physics
scenarios, as they are the ubiquitous prediction of spontaneous Peccei-Quinn
(PQ)~\cite{Peccei:1977hh} symmetry breaking.  The most famous example is the
Weinberg-Wilczek axion invoked to solve the strong $CP$ problem
\cite{Weinberg:1977ma,Wilczek:1977pj}, as well as invisible axion variants
\cite{Dine:1981rt,Zhitnitsky:1980tq,Kim:1979if,Shifman:1979if}.  Light
pseudoscalar particles appear in any Higgs sector with an approximate PQ
symmetry, which often occurs in the minimal or next-to-minimal supersymmetric
standard models (MSSM and NMSSM).  Models of dynamical supersymmetry breaking
typically predict an $R$-axion~\cite{Nelson:1993nf}, whose couplings can mimic
PQ-type axions.  There has also been speculation~\cite{He:2006uu} that the
HyperCP anomaly~\cite{Park:2005eka} might be explained by a light axion. 
Therefore, searches for light axionlike states have the potential to confirm or
exclude a variety of new physics models. 

In this paper, we show that flavor-changing neutral current $b \rightarrow s$
decays place stringent bounds on such models.  While the coupling of the axion
to fermions is flavor-diagonal, the $b \rightarrow s a$ decay mediated by a
top-$W$ penguin diagram is enhanced by the top Yukawa coupling appearing in the
top-axion vertex.  To our knowledge,
Refs.~\cite{Wise:1980ux,Hall:1981bc,Frere:1981cc} were the first to consider
this decay as a search channel for light pseudoscalars, where the $a$ field was
identified with the $CP$-odd Higgs $A^0$ in a two Higgs doublet model (2HDM). 
The goal of this paper is to revive this search channel in models like the axion
portal, where there is an $a$ field which mixes with $A^0$.

In the parameter range of interest for the axion portal, the axion decays
promptly to $\mu^+ \mu^-$, and we find that existing $B\to K\ell^+\ell^-$ data
(for $\ell = e,\mu$) can be used to derive multi-TeV constraints on the axion
decay constant $f_a$, especially for small values of $\tan \beta$. For heavier
axionlike states with reduced branching fractions to muons, $B\to
K\ell^+\ell^-$ can still be used to place a bound, relevant for constraining
light Higgs scenarios in the
NMSSM~\cite{Dermisek:2005ar,Chang:2008cw,Dermisek:2008uu}. The estimates in this
paper are likely improvable by dedicated $B \rightarrow K a$ searches at BaBar
and Belle, and can be further strengthened at LHCb and a possible
super $B$-factory. These searches are complementary to $\Upsilon(nS) \rightarrow
\gamma a$ searches recently performed by BaBar~\cite{Aubert:2009cp}.  

In the next section, we review the axion portal Lagrangian, which is relevant
for any DFSZ-type (Dine-Fischler-Srednicki-Zhitnitsky) axion~\cite{Dine:1981rt,Zhitnitsky:1980tq}, and use it to
calculate the effective $b\to sa$ vertex in \Sec{sec:vert}.  We sketch the
current experimental situation in \Sec{sec:branch} and derive corresponding
bounds in \Sec{sec:interpret}. We conclude in \Sec{sec:conc}.

\section{Review of the Axion Portal}
\label{sec:model}

If one were only interested in studying the tree-level interactions of new
axionlike states, it would be sufficient to introduce a new term in the
Lagrangian of the form
\be
\label{eq:simplecouping}
\delta\mathcal{L}_{\text{int}} = \frac{c_\psi}{f_a}\, 
  \overline{\psi}\gamma_\mu\gamma_5\psi\, \partial_\mu a \,,
\ee
where $f_a$ is the axion decay constant and $c_\psi$ is the fermion charge under
the broken $U(1)$.  By the equations of motion, such a coupling is proportional
to the fermion mass parameter, leading to an effective coupling constant $c_\psi
m_\psi / f_a$.  However, the $b \rightarrow s a$ process we are interested in
occurs via a top-$W$ penguin loop.  With only \Eq{eq:simplecouping}, such a
diagram is logarithmically sensitive to the cutoff scale~\cite{Wise:1980ux}, so
it is necessary to embed the axion coupling in a complete theory to get a
reliable bound on $f_a$.

The axion portal~\cite{Nomura:2008ru} is an example of a class of theories where
the $b \rightarrow s a$ amplitude is finite. The axion arises from
spontaneous PQ-symmetry breaking in a 2HDM, of which the DFSZ axion is a special
case. We show that the $b \rightarrow s a$ amplitude can be derived from the
$b \rightarrow s A^0$ amplitude, where $A^0$ is the $CP$-odd Higgs boson
in a PQ-symmetric 2HDM.

Consider a complex scalar field $S$ carrying $U(1)_{\rm PQ}$ charge that gets a
vacuum expectation value $\langle S \rangle \equiv f_a$.  This spontaneous
symmetry breaking leads to a light axionlike state, $a$,
\be
\label{eq:Sdef}
S = f_a \exp\left[\frac{i}{\sqrt{2} f_a}\, a \right].
\ee
The assumption in the axion portal (and for any DFSZ-type axion) is that the
only operator that transmits PQ charge from $S$ to the standard model is
\be
\label{eq:DFSZmass}
\delta \mathcal{L} =  \lambda\, S^n h_u h_d + \text{h.c.},
\ee
where $\lambda$ is a (possibly dimensionful) parameter, and $n$ is an integer. 
This coupling forces $h_u h_d$ to carry nontrivial PQ charge, and we assume
that the entire Higgs potential preserves the $U(1)_{\rm PQ}$ symmetry to a good
approximation. The DFSZ axion~\cite{Dine:1981rt,Zhitnitsky:1980tq} corresponds
to the case with $n=2$, while for the PQ-symmetric NMSSM~\cite{Hall:2004qd} $n = 1$.
Either case can be used in the axion portal model of dark matter.  

Since the vacuum expectation values (VEVs) of $S$, $h_u$, and $h_d$ all break
the PQ symmetry, the physical axion will be a linear combination of the phases
of all three fields.\footnote{This also means that the physical axion decay
constant will be a function of the three VEVs.  The difference is negligible
when $f_a \gg v_{\rm EW}$, and we will continue to refer to $f_a$ as the axion
decay constant.}  However, in the $f_a \gg v_{\rm EW}$ limit, it is
calculationally more convenient to work in an ``interaction eigenstate'' basis,
where the axion $a$ only appears in $S$, and the $CP$-odd Higgs $A^0$ only
appears in the two Higgs doublets in the form:
\beqa\label{eq:exppara}
h_u &=& \left( \begin{array}{c} \displaystyle
  v_u\, \exp \left[\frac{i \cot \beta}{\sqrt{2} v_{\rm EW}}\, A^0\right]\\[8pt]
  0 \end{array} \right) , \nn\\
h_d &=& \left( \begin{array}{c}  0 \\[2pt] \displaystyle
  v_d\, \exp \left[\frac{i \tan \beta}{\sqrt{2} v_{\rm EW}}\, A^0 \right] 
  \end{array} \right),
\eeqa
where $\tan \beta \equiv {v_u}/{v_d}$,
\be
v_{\rm EW} \equiv \sqrt{v_u^2 + v_d^2} \equiv \frac{m_W}{g} 
  \simeq 174 \GeV\,,
\ee
and we have omitted the charged Higgs $H^\pm$ and the $CP$-even Higgses for
simplicity.  The coefficients appearing in front of $A^0$ ensure that $A^0$ is
orthogonal to the Goldstone boson eaten by the $Z$ boson.

This exponential parameterization of $A^0$ is convenient for our purposes, since
PQ symmetry implies that mass terms involving $a$ and $A^0$ can only appear in
\Eq{eq:DFSZmass}.  In this basis, the physical degrees of freedom are given by
\beqa\label{eq:mixangle}
a_{\rm phys.} &=& a \cos \theta  - A^0 \sin \theta, \nn\\
A^0_{\rm phys.} &=& a \sin \theta  + A^0 \cos \theta,
\eeqa
with
\be\label{thetadef}
\tan \theta \equiv n\, \frac{v_{\text{EW}}}{f_a}\, \frac{\sin 2\beta}{2}\,.
\ee
At this level, the physical axion is massless.\footnote{For completeness, the
physical $A^0$ mass is given by $m^2({A^0_{\rm phys.}}) = \lambda\, (f_a)^n\, (2
/ \sin 2\beta) (1 + \tan^2 \theta)$.}  A small mass (beyond the contribution
from the QCD anomaly) can be generated by a small explicit violation of the PQ
symmetry, but the precise way this happens is irrelevant for our discussion.

The dominant decay mode for the axion depends on its mass, $m_a$.  The axion
decay width to an $\ell^+\ell^-$ lepton pair is given by 
\be
\label{eq:Br2leptons}
\Gamma(a \rightarrow \ell^+ \ell^-) = n^2 \sin^4 \beta\, 
  \frac{m_a}{16 \pi}\, \frac{m_{\ell}^2}{f_a^2}\, 
  \sqrt{1-\frac{4m_\ell^2}{m_a^2}} \,.
\ee
For $2m_e < m_a < 2m_\mu$, the dominant decay is $a\to e^+e^-$.  In this mass
range, however, strong bounds already exist from $K \to \pi a$
decays~\cite{Anisimovsky:2004hr,Adler:2001xv}.  With the axion decay to fermions
being proportional to their mass-squared, $a \rightarrow \mu^+\mu^-$ dominates
over $a \rightarrow e^+e^-$ for $m_a > 2m_\mu$.  Note that in the mass range
given in \Eq{jyrange}, the axion decays within the detector as long as $f_a
\lesssim 1000 \TeV$.

The axion decay becomes more complicated at higher masses when hadronic decay
modes open up.  Reference~\cite{Nomura:2008ru} estimated that the $a
\rightarrow 3 \pi$ channel starts to dominate over the $\mu^+ \mu^-$ channel at
$m_a \simeq 800\MeV$.  Hadronic channels dominate the axion decay until $m_a
\gtrsim 2m_\tau$, when the $\tau^+\tau^-$ channel becomes dominant.  However, as
emphasized recently in~\cite{Lisanti:2009uy}, throughout the entire mass range
$2 m_\mu < m_a < 2 m_b$, the branching ratio to $\mu^+\mu^-$ remains
significant, and until the $\tau^+\tau^-$ threshold, it never drops below
$\mathcal{O}(10^{-2})$.  For $m_a > 2m_\tau$, the branching fraction to muons is
approximately
\be
\label{eq:brtotaus}
\Br(a \rightarrow \mu^+ \mu^-) \simeq \frac{m_\mu^2}{m_\tau^2} \simeq 0.003,
\ee
with the precise value depending on $\tan \beta$ through $\Gamma(a \rightarrow c
\bar{c})$ and on the neglected phase space factor.

\section{\boldmath The Effective $b\to sa$ Coupling}
\label{sec:vert}

By assumption, the physical axion state dominantly couples to standard model
fields via its mixing with $A^0$.  Therefore, at one-loop level, the amplitude
for $b \rightarrow s a$ can be derived from
\be\label{PQrel1}
\mathcal{M}(b \to s a)
  = - \sin \theta \times \mathcal{M}(b \to s A^0)_{\rm 2HDM},
\ee
where ``2HDM'' refers to a (PQ-symmetric) 2HDM with no $S$ field.  Moreover,
since the final state only contains a single axion field, there is no difference
in the relevant Feynman rules between the exponential parameterization in
\Eq{eq:exppara} and the standard linear parameterization of $A^0$ in the two
Higgs doublet literature.  For concreteness, we will consider a type-II
(MSSM-like) 2HDM.\footnote{The type-I 2HDM model gives the same $b \rightarrow s
A^0$ amplitude to the order we are working; see Ref.~\cite{Hall:1981bc}.}

The radiatively induced $b \to sA^0$ coupling in a type-II 2HDM was calculated
in the early 1980's independently in two
papers~\cite{Hall:1981bc,Frere:1981cc}.  The dominant contributions come from
penguin diagrams involving a top quark, a $W$ boson and/or charged Higgs $H^\pm$
boson, and the $t\,\overline t A^0$ or $W^\pm H^\mp A^0$ couplings (and
corresponding counterterms).  The one-loop $b\to sA^0$ amplitude is reproduced
to lowest order (in the $m_{B,A^0} \ll m_{W,t,H}$ limit) by the tree-level
matrix element of the effective
Hamiltonian~\cite{Hall:1981bc,Frere:1981cc}\footnote{The results published in
these two papers differ, a fact which seems to have gone unnoticed---or at least
unremarked upon---in the literature.  We have redone the calculations both in
the unitary gauge and in the Feynman gauge and agree with the result in
Ref.~\cite{Hall:1981bc}.  We also agree with Ref.~\cite{Frere:1981cc} if we
replace in their Eq.~(9) the second $\ln(m_t^2/m_W^2)$ term by
$\ln(m_t^2/m_H^2)$, most likely indicating a simple typographical error. 
Several papers in the literature  seem to use the result as printed in
Ref.~\cite{Frere:1981cc}, which has qualitatively wrong implications.  For
example, it exhibits decoupling in the $m_H\to \infty$ limit and singularities
when $m_H \rightarrow m_t$, whereas the correct result does not.}
\be\label{Heff}
\mathcal{H} = \frac{g^3\, V_{ts}^* V_{tb}}{128\,\pi^2}\,
  \frac{m_t^2}{m_W^3} \left(X_1\cot\beta + X_2\cot^3\beta\right)
  \bar s\gamma^\mu P_L b\, \partial_\mu A^0 .
\ee
The functions $X_1$ and $X_2$ depend on the charged Higgs boson mass $m_H$, and
are given by
\beqa
X_1 &=& 2 + \frac{m_H^2}{m_H^2 - m_t^2} - \frac{3m_W^2}{m_t^2 - m_W^2} \nn\\
&&{} + \frac{3 m_W^4 (m_H^2+m_W^2-2m_t^2)}{(m_H^2-m_W^2)\, (m_t^2-m_W^2)^2}\, 
  \ln\frac{m_t^2}{m_W^2} \nn\\
&&{} + \frac{m_H^2}{m_H^2 - m_t^2}\left( \frac{m_H^2}{m_H^2 - m_t^2}
  - \frac{6m_W^2}{m_H^2 - m_W^2} \right)\ln\frac{m_t^2}{m_H^2} \,, \nn\\
X_2 &=& - \frac{2m_t^2}{m_H^2 - m_t^2}
  \left( 1 + \frac{m_H^2}{m_H^2 - m_t^2}\, \ln\frac{m_t^2}{m_H^2} \right) .
\eeqa

From this effective Hamiltonian, we can calculate various $B$ decay rates in the
2HDM.  These are summarized in the Appendix for $B \rightarrow K a$, $B
\rightarrow K^* a$, and the inclusive $B \rightarrow X_s a$ rates.  Using
\Eq{PQrel1}, the rates in any of these channels relevant for the axion portal
are determined by
\be\label{PQrel2}
\Gamma(B \to K a)
  = \sin^2 \theta \times \Gamma(B \to K A^0)_{\rm 2HDM}.
\ee

\section{Experimental Bounds}
\label{sec:branch}

In the parameter range of interest, the axion has a significant decay rate to
leptons and decays promptly on collider timescales. Thus, the axion would
manifest itself as a narrow dilepton peak in $b \rightarrow s\ell^+\ell^-$
decays.  

The $b\to s a \to s\,\ell^+\ell^-$ process contributes to both inclusive and
exclusive $B\to X_s\ell^+\ell^-$ decays~\cite{Ali:1999mm,Hiller:2004ii}.  These
final states receive large long-distance contributions from intermediate
$J/\psi$ and $\psi'$ resonances decaying to $\ell^+\ell^-$, which result in
removing the surrounding $q^2 (\equiv m^2_{\ell^+\ell^-})$ regions from the
measurements.  The so-called low-$q^2$ region ($q^2 \lesssim 7-8\GeV^2$) can
probe axion masses up to $m_a \sim 2.7\GeV$, while the high-$q^2$ region ($q^2
\gtrsim 14\GeV^2$) is above the $a\to \tau^+\tau^-$ threshold.  In general, one
can bound the axion contribution in both these regions.

In the low-$q^2$ region, and especially for $m_a \lesssim 800$~MeV as in
Eq.~(\ref{jyrange}), the exclusive mode $B\to K\ell^+\ell^-$ is particularly
well-suited to constrain $b\to sa$.  This is because $\d\Gamma(B\to
K\ell^+\ell^-)/\d q^2$  varies slowly at small $q^2$, and $B\to K\ell^+\ell^-$
has a smaller rate than $B\to K^*\ell^+\ell^-$, thus it gives us the best bound
by simply looking at the measured spectrum. In contrast,  the exclusive $B\to
K^*\ell^+\ell^-$ and the inclusive $B\to X_s\ell^+\ell^-$ decay modes receive
large enhancements from the electromagnetic penguin operator, whose contribution
rises steeply at small $q^2$, as $1/q^2$.  This will complicate looking for a
small excess in these modes in this region.

For $m_a \gtrsim 1$~GeV, we expect that the bounds from $B\to K\ell^+\ell^-$ and
$K^*\ell^+\ell^-$ may be comparable (possibly even from $B\to X_s\ell^+\ell^-$
if a super $B$-factory is constructed), and a dedicated experimental analysis
should explore how to set the strongest bound, using the rate predictions in
App.~\ref{sec:decayrates}.  For the remainder of this paper, we focus on $B\to
K\ell^+\ell^-$.

Since $B\to Ka$ contributes mostly to the $K\mu^+\mu^-$ final state, and much
less to $Ke^+e^-$, to set the best possible bound on $B\to Ka$, one needs the
$B\to K\mu^+\mu^-$ and $B\to Ke^+e^-$ spectra separately.  This information does
not seem to be available in the published
papers~\cite{Wei:2009zv,Aubert:2008ps}.  Based on the latest world average,
$\Br(B\to K\ell^+\ell^-) = (4.5 \pm 0.4) \times
10^{-7}$~\cite{Barberio:2008fa,Wei:2009zv,Aubert:2008ps}, and the spectrum in
Fig.~1 in Ref.~\cite{Wei:2009zv}, it seems to us that
\be\label{guessedbound}
\Br(B\to Ka) \times \Br(a\to \mu^+\mu^-) < 10^{-7}
\ee
is a conservative upper bound for any value of the axion mass satisfying
$m_a<m_B-m_K$.

As we emphasized, BaBar, Belle, and a possible super $B$-factory should be able
to set a better bound on a narrow resonance contributing to $B\to
K^{(*)}\mu^+\mu^-$ but not to $B\to K^{(*)}e^+e^-$.  Moreover, LHCb will also be
able to search for deviations from the standard model predictions in $B\to
K^{(*)} \ell^+\ell^-$ with significantly improved sensitivity.  While we could
not find a recent LHCb study for the $K \ell^+\ell^-$ mode (only for $K^*
\ell^+\ell^-$), the fact that the signal to background ratio at the $e^+e^-$
$B$-factories is not worse in $B\to K \ell^+\ell^-$ than in $B\to K^*
\ell^+\ell^-$ suggests that LHCb should be able to carry out a precise
measurement~\cite{Uli}.  Interestingly, since the $B\to Ka$ signal is
essentially a delta function in $q^2$, the bound in Eq.~(\ref{guessedbound}) can
be improved as experimental statistics increase by considering smaller and
smaller bin sizes, without being limited by theoretical uncertainties in form
factors~\cite{Ball:2004ye} (or by nonperturbative
contributions~\cite{Bartsch:2009qp}).  The bound on $f_a$ will increase compared
to the results we obtain in the next section, simply by scaling with the bound
on $1/\sqrt{\Br(B\to Ka)}$.

\section{Interpretation}
\label{sec:interpret}

\begin{figure}[tb]
\includegraphics[width=\columnwidth]{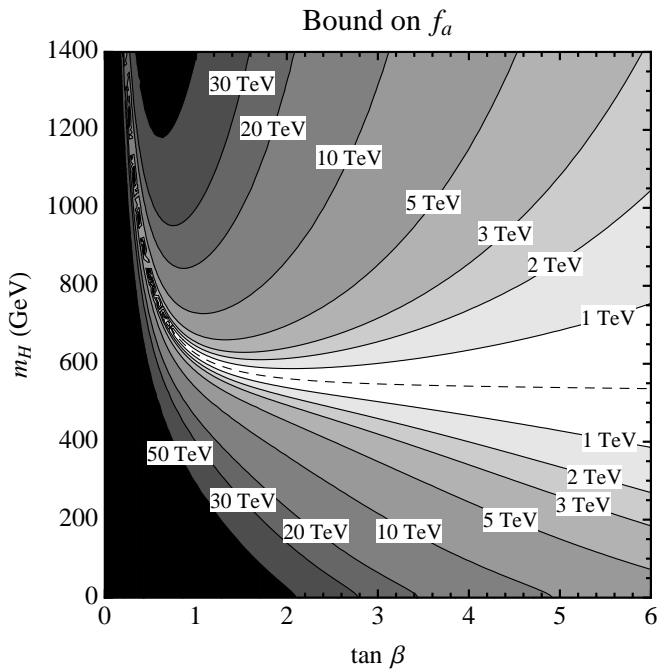}
\caption{Bounds on $f_a$ as a function of $\tan\beta$ and $m_H$ for $n=1$ in
Eq.~(\ref{thetadef}), for $m_a^2\ll m_B^2$.  For each displayed value of $f_a$
there are two contour lines, and the region between them is allowed for $f_a$
below the shown value. The bound disappears along the dashed curve, and gets
generically weaker for larger $\tan\beta$.}
\label{fig:fabound}
\end{figure}

We now derive the bounds on $f_a$ using the calculated $B \rightarrow K a$
branching ratio in \Eq{PQrel2} and the experimental bound in \Eq{guessedbound}. 
We start with the axion portal scenario with $\Br(a \rightarrow \mu^+ \mu^-)
\sim 100\%$ and where $\sin \theta$ is defined in terms of $f_a$ by
\Eq{thetadef}.  We will then look at the bound on more general scenarios,
including the light Higgs scenario in the NMSSM.

For the axion portal, \Fig{fig:fabound} shows the constraints on $f_a$ as a
function of the charged Higgs boson mass $m_H$ and $\tan \beta$.  For
concreteness, we take $n=1$; other values of $n$ correspond to a trivial scaling
of $f_a$.  In the mass range in Eq.~(\ref{jyrange}), the dependence on $m_a$ is
negligible for setting a bound.  The bound on $f_a$ is in the multi-TeV range
for low values of $\tan \beta$ and weakens as $\tan \beta$ increases.  At each
value of $\tan\beta$, there is a value of $m_H$ for which the $b\to s a$
amplitude in \Eq{Heff} changes signs, indicated by the dashed curve in
\Fig{fig:fabound}, along which the bound disappears.  Higher order corrections
will affect where this cancellation takes place, but away from a very narrow
region near this dashed curve, the derived bound is robust. 
The region $\tan\beta < 1$ is constrained by the top Yukawa
coupling becoming increasingly nonpertubative; this region is included in
Figs.~\ref{fig:fabound} and \ref{fig:st2bound}, nevertheless, to provide a
clearer illustration of the parametric dependence of the bounds.

\begin{figure}[tb]
\includegraphics[width=\columnwidth]{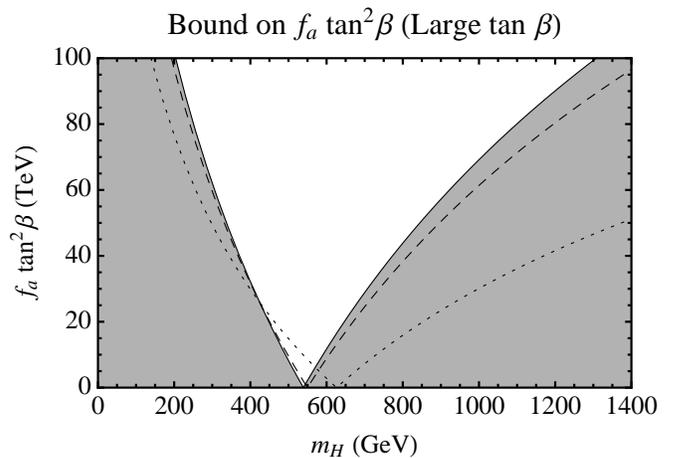}
\caption{The shaded regions of $f_a \tan^2 \beta$ are excluded in the large
$\tan \beta$ limit.  To indicate the region of validity of the large $\tan
\beta$ approximation, the dashed (dotted) curve shows the bound for $\tan \beta
= 3$ ($\tan \beta = 1$).}
\label{fig:fabound_largetb}
\end{figure}

\begin{figure}[tb]
\includegraphics[width=\columnwidth]{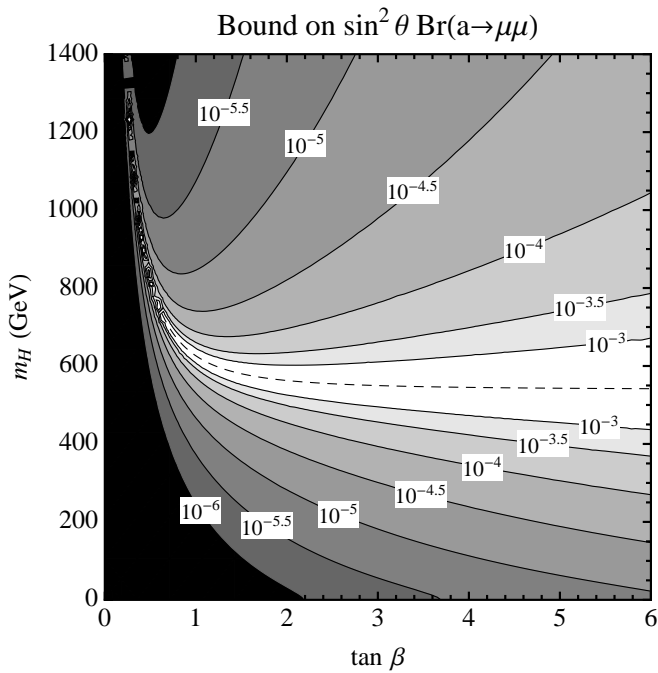}
\caption{Bounds on $\sin^2\theta\, \Br(a\to\mu^+\mu^-)$ as a function of
$\tan\beta$ and $m_H$.  Similar to Fig.~\ref{fig:fabound}, the successively
darker regions going away from the dashed curve are allowed for $\sin^2\theta\,
\Br(a\to\mu^+\mu^-)$ above the indicated values.  When $m_a$ is not small
compared to $m_B$, these bounds should be modified by Eq.~(\ref{abovetau}), but
this is a small effect.}
\label{fig:st2bound}
\end{figure}

As one goes to large values of $\tan \beta$, the $X_1$ piece of \Eq{Heff}
dominates, and $\sin (2\beta) / 2 = 1/\tan \beta + {\cal O}(1/\tan^3\beta)$.  In
this limit, the constraint takes a particularly simple form that only depends on
the combination $f_a \tan^2 \beta$, as shown in \Fig{fig:fabound_largetb}. 
Except in the region close to $m_H \sim 550 \GeV$, the bound is better than $f_a
\tan^2 \beta \gtrsim \text{few} \times 10 \TeV$.

These $B \rightarrow Ka$ bounds are complementary to those recently set by
BaBar~\cite{Aubert:2009cp} in $\Upsilon(nS) \to \gamma\, a \to \gamma\, 
\mu^+ \mu^-$:
\be
f_a \gtrsim  (1.4 \TeV) \times \sin^2 \beta \,.
\ee
For example, for $m_H \simeq 400 \GeV$, the $\Upsilon$ bound dominates for $\tan
\beta \gtrsim 5$, while $B \rightarrow Ka$ dominates for $\tan \beta \lesssim
5$.

The bounds in Figs.~\ref{fig:fabound} and \ref{fig:fabound_largetb} apply for a
generic axion portal model where $m_H$ and $\tan \beta$ are free parameters. 
One would like some sense of what the expected values of $m_H$ and $\tan \beta$
might be in a realistic model.  Ref.~\cite{Nomura:2008ru} considered a specific
scenario based on the PQ-symmetric NMSSM~\cite{Hall:2004qd}.  In that model
small $\tan \beta$ is preferred, since large $\tan \beta$ requires fine-tuning
the Higgs potential.  In addition, $m_H$ is no longer a free parameter and is
approximately related to the mass of the lightest $CP$-even scalar $s_0$ via
\be
m_H^2 \simeq m_W^2 + \left(\frac2{\sin^2 2\beta}\, 
  \frac{m_{s_0} f_a}{v_{\rm EW}} \right)^2.
\ee
In the context of dark matter, Ref.~\cite{Nomura:2008ru} required $m_{s_0}$ to
be $\mathcal{O}(10 \GeV)$ to achieve a Sommerfeld enhancement.  Taking $m_{s_0}
= 10 \GeV$ and $f_a = 2 \TeV$ as a benchmark, the $B \rightarrow K a$ bound
requires $2.5 \lesssim \tan \beta \lesssim 3.0$, corresponding to $490\GeV
\lesssim m_H \lesssim 650 \GeV$, in the vicinity of the cancellation region. 
This bound is very sensitive to $m_{s_0}$; for $m_{s_0} = 20 \GeV$ and $f_a = 2
\TeV$, the bounds are $1.5 \lesssim \tan \beta \lesssim 1.7$ and $550\GeV
\lesssim m_H \lesssim 610 \GeV$.  Note that models like~\cite{Mardon:2009gw}
have no preferred value of $m_H$, can have larger values of $f_a$, and do not
disfavor large $\tan \beta$.

As mentioned, these $B \rightarrow K a$ constraints apply to any scenario where
the branching ratio formula in \Eq{PQrel2} applies, i.e.\ where the axion
couplings are determined via \Eq{eq:mixangle}, and where $m_a < m_B  - m_K$. 
For example, recent studies of light Higgs bosons in the
NMSSM~\cite{Dermisek:2005ar,Chang:2008cw,Dermisek:2008uu} and related dark
matter constructions \cite{Bai:2009ka,Hooper:2009gm} also contain a light
pseudoscalar whose couplings to standard model fermions can be described in
terms of a mixing angle $\theta$, as in Eq.~(\ref{thetadef}).\footnote{In the
literature, $\sin \theta$ is often referred to as the ``non-singlet fraction''
$\cos \theta_A$~\cite{Chang:2008cw}.}  There, the mass of the $a$ field is
expected to be $2 m_\tau < m_a < 2 m_b$, with the $a \rightarrow \mu^+ \mu^-$
branching fraction estimated in \Eq{eq:brtotaus}.

\begin{figure}[tb]
\includegraphics[width=\columnwidth]{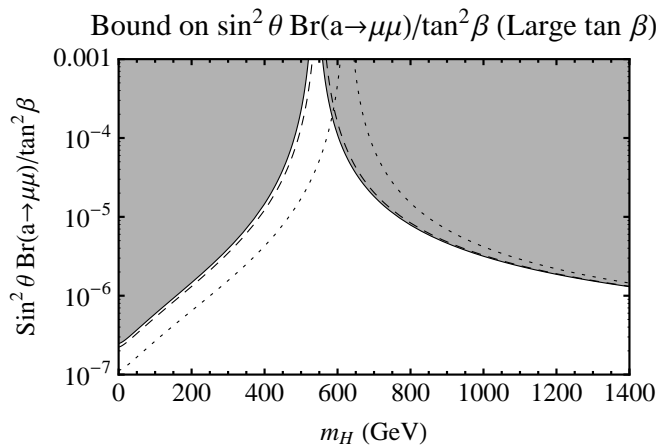}
\caption{Bounds on $\sin^2\theta\, \Br(a\to\mu^+\mu^-)/\tan^2 \beta$ in the
large $\tan \beta$ limit.  The shaded region is excluded, and the dashed
(dotted) curve shows $\tan \beta = 3$ ($\tan \beta = 1$).}
\label{fig:st2bound_largetb}
\end{figure}

To show the constraints on such scenarios in a model independent way, we plot
the bound on the combination $\sin^2\theta\, \Br(a\to\mu^+\mu^-)$ in
\Fig{fig:st2bound}, in the $m_a^2 \ll m_B^2$ limit for simplicity.  We also show
the large $\tan \beta$ limit in \Fig{fig:st2bound_largetb}, where the bound is
on the combination $\sin^2\theta\, \Br(a\to\mu^+\mu^-)/\tan^2 \beta$.  To apply
these bounds for the case where $m_a$ is not small compared to $m_B$, one should
make the replacement in Figs.~\ref{fig:st2bound} and \ref{fig:st2bound_largetb}
(see the Appendix),
\be\label{abovetau}
\sin^2 \theta \,\Rightarrow\, \sin^2\theta\,
  \frac{\lambda_K(m_a)\, \big[f_0(m_a^2)\big]^2}
  {(m_B^2-m_K^2) \big[f_0(0)\big]^2}
\equiv \sin^2\theta\, R(m_a)\,.
\ee
Using a simple pole form for the $q^2$ dependence of $f_0$~\cite{Ball:2004ye},
we find that $R(m_a)$ deviates from unity by less than 20\% for $m_a < 4.6$~GeV
(i.e.\ nearly over the full kinematically allowed region), and so it is a good
approximation to neglect $R(m_a)$.  In the case of NMSSM scenarios, the precise
bound depends strongly on the parameters of the theory.  To give a sense of the
strength of the bound, for $m_a \sim 4\GeV$, $m_H \sim 200 \GeV$, and using
Eq.~(\ref{eq:brtotaus}), the bound at large $\tan\beta$ implies $\sin^2 \theta /
\tan^2 \beta \lesssim 5\times 10^{-4}$ (and $\sin^2 \theta \lesssim 2\times
10^{-4}$ for $\tan\beta=1$), which is a significant constraint on large mixing
angles or small~$\tan \beta$.

\section{Conclusions}
\label{sec:conc}

In this paper, we explored bounds on axionlike states from flavor-changing
neutral current $b\to s$ decays.  We found that the exclusive $B\to
K\ell^+\ell^-$ decay is particularly well-suited to constrain such
contributions.  In the case of the axion portal (or equivalently, any DFSZ-type
axion), we derived a bound from current $B$-factory data on the axion decay
constant $f_a$.  The bound is in the multi-TeV range, gets stronger for small
$\tan \beta$, and depends sensitively on the value of the charged Higgs boson
mass.  This places tension on the axion portal model of dark matter in the
parameter space given in \Eq{jyrange}.  More generally, there is a constraint on
any pseudoscalar with $2m_\mu < m_a < m_B - m_K$ whose couplings to standard
model fermions arise via mixing with the $CP$-odd Higgs $A^0$.  This is true
even if $\Br(a \rightarrow \mu^+ \mu^-) \sim \mathcal{O}(10^{-3})$, as is the
case for light Higgs scenarios in the NMSSM.

We derived our bound using a conservative estimate from the $q^2$ distribution
in $B\to K\ell^+\ell^-$.  The bound could most probably be improved through a
dedicated search in existing $B$-factory data, and in searches at LHCb and a
possible future super $B$-factory.  The $B \rightarrow Ka$ search is
complementary to axion searches in $\Upsilon(nS) \rightarrow \gamma a$, because
for fixed mixing angle $\theta$ in a type-II 2HDM, the former scales like
$1/\tan^2 \beta$ while the latter scales like $\tan^2 \beta$.

One way to extend our analysis would be to look at axions decaying to hadronic
final states.  We focused on the decay mode $a \rightarrow \mu^+ \mu^-$, since
the $a \rightarrow e^+ e^-$ mode is already well-constrained by kaon decays, and
we were motivated by the parameter space relevant for
Ref.~\cite{Nomura:2008ru}.  However, as the axion mass increases, other decay
channels open up, such as $a\to\pi^+\pi^-\pi^0$, $a\to K K^{*}$, etc.  These
would also be worthwhile to search for in $B$-factory data, especially since
dark matter models such as~\cite{Mardon:2009gw} are compatible with
$a\to\pi^+\pi^-\pi^0$ decays.  It appears to us that setting bounds in these
modes is more complicated than for $B\to K\ell^+\ell^-$, and should be done in
dedicated experimental analyses.  For constraining higher mass axions, it would
be interesting to study whether $B$-factories could search for narrow resonances
in $B\to K\tau^+\tau^-$ at a level of sensitivity no weaker than
$m_\tau^2/m_\mu^2$ times the corresponding bound in $B\to K\mu^+\mu^-$. 
Combining a number of search channels, one would be able to substantially probe
scenarios containing light axion-like states.

\bigskip
\textbf{Note Added}: While this paper was being completed, \Ref{Batell:2009jf}
appeared, which claims much stronger bounds on $f_a$ than our result.  They use
a  different effective Hamiltonian from \Eq{Heff}, which does not include the
effect of charged Higgs bosons, crucial for bounding DFSZ-type axions.

\begin{acknowledgments}

This paper was inspired by a talk by Maxim Pospelov at the SLAC Dark Forces
workshop in September 2009.  We thank Lawrence Hall and Mark Wise for helpful
conversations, and we apologize for making them (temporarily) worried about the
correctness of their result.  We benefitted from the advice of Mariangela
Lisanti, Yasunori Nomura, and Jay Wacker.  This work was supported in part by
the Director, Office of Science, Office of High Energy Physics of the U.S.\
Department of Energy under Contract No.~DE-AC02-05CH11231.

\end{acknowledgments}

\bigskip\bigskip\bigskip
\appendix*

\section{Decay Rates}
\label{sec:decayrates}

In this Appendix, we list the $B$ decay rates to $K^{(*)} A^0$ and $X_s A^0$ in
the 2HDM, using the effective Hamiltonian in \Eq{Heff}.  These should be
combined with Eq.~(\ref{PQrel2}) to bound the axion models.

Defining
\be\label{G0def}
\Gamma_0 = \frac{G_F^3 |V_{ts}^* V_{tb}|^2}{\sqrt2\, 2^{12}\, \pi^5}\, m_t^4\,
  m_B^3 \big(X_1\cot\beta + X_2\cot^3\beta\big)^2 ,
\ee
and
\be
\lambda_{K^{(*)}} = \sqrt{\big(m_B^2-m_{A^0}^2-m_{K^{(*)}}^2\big)^2 
  - 4 m_{A^0}^2 m_{K^{(*)}}^2} \,,
\vspace*{10pt}
\ee
the $B\to KA^0$ decay rate is given by
\be\label{BKarate}
\Gamma(B\to KA^0) = \Gamma_0\,
  \frac{\lambda_K (m_B^2-m_K^2)^2}{m_B^6}\, \big[f_0(m_{A^0}^2)\big]^2 .
\ee
The $B\to K^*a$ decay rate is
\be\label{BKsarate}
\Gamma(B\to K^*\!A^0) = \Gamma_0\, \frac{\lambda_{K^*}^3}{m_B^6}\,
  \big[A_0(m_{A^0}^2)\big]^2 .
\ee
In both decays we used the standard definitions~\cite{Ali:1999mm} of the form
factors,
\beqa\label{BKff}
\langle K(p-q)| \bar s\, q\!\!\!\slash P_L b\, |
  B(p)\rangle &=& \frac12 (m_B^2 - m_K^2)\, f_0(q^2)\,,\qquad\quad\\
\langle K^*(p-q)| \bar s\, q\!\!\!\slash P_L b\, |
  B(p)\rangle &=& -i\, m_{K^*} (\varepsilon^*\cdot p)\, A_0(q^2)\,. \nn
\eeqa
(We caution the reader not to confuse $A^0$ and $A_0$, each of which are
standard in the respective contexts.)

In Eq.~(\ref{G0def}) it is the $\overline{\text{MS}}$ top quark mass which
enters, appropriate both for the coupling to Higgses and in loop integrals. 
While this distinction is formally a higher order correction, since the rates
are proportional to $m_t^4$, we use the Tevatron average top mass, converted to
$\overline{\text{MS}}$ at one-loop, $\overline m_t = m_t [1 - 4\alpha_s/(3\pi)]
\approx 165$~GeV.

The largest hadronic uncertainty in evaluating the implication of the bound in
Eq.~(\ref{guessedbound}) is the model dependence in the calculations of the form
factor $f_0(m_a^2)$, which is an increasing function of $q^2$.  For $f_0(0)$,
QCD sum rule calculations obtain values around 0.33, with an order $10\%$
uncertainty~\cite{Ball:2004ye}.  To be conservative, in evaluating the bound on
$f_a$, we only assume $f_0(0) > 0.25$ for $m_a\ll m_B$ (which also covers lower
values motivated by the fit in Ref.~\cite{Arnesen:2005ez}).  For $m_a \gtrsim
2m_\tau$, relevant for Eq.~(\ref{abovetau}), we use the approximation $f_0(q^2)
= f_0(0) / (1-q^2/37.5\GeV^2)$~\cite{Ball:2004ye}, which should be good enough
for our purposes. For recent QCD sum rule calculations of $A_0(q^2)$, relevant
for setting a bound using $B\to K^*\ell^+\ell^-$, see Ref.~\cite{Ball:2004rg}.

The inclusive $B\to X_s\, a$ decay rate, which can be calculated (strong
interaction) model independently in an operator product expansion, is given at
leading order in $\Lambda_{\rm QCD}/m_b$ by
\be\label{BXsarate}
\Gamma(B\to X_s A^0) = 2 \Gamma_0\, \frac{m_b^3}{m_B^3}
  \bigg(1-\frac{m_{A^0}^2}{m_b^2}\bigg)\,. \\
\ee

\end{document}